\documentclass[12pt,epsf]{article}
\usepackage{amsmath}
\usepackage{amssymb}
\begin{document}
\def\a{\alpha}
\def\b{\beta}
\def\c{\varepsilon}
\def\d{\delta}
\def\e{\epsilon}
\def\f{\phi}
\def\g{\gamma}
\def\h{\theta}
\def\k{\kappa}
\def\l{\lambda}
\def\m{\mu}
\def\n{\nu}
\def\p{\psi}
\def\q{\partial}
\def\r{\rho}
\def\s{\sigma}
\def\t{\tau}
\def\u{\upsilon}
\def\v{\varphi}
\def\w{\omega}
\def\x{\xi}
\def\y{\eta}
\def\z{\zeta}
\def\D{{\mit \Delta}}
\def\G{\Gamma}
\def\H{\Theta}
\def\L{\Lambda}
\def\F{\Phi}
\def\P{\Psi}

\def\S{\Sigma}

\def\o{\over}
\def\beq{\begin{eqnarray}}
\def\eeq{\end{eqnarray}}
\newcommand{\gsim}{ \mathop{}_{\textstyle \sim}^{\textstyle >} }
\newcommand{\lsim}{ \mathop{}_{\textstyle \sim}^{\textstyle <} }
\newcommand{\vev}[1]{ \left\langle {#1} \right\rangle }
\newcommand{\bra}[1]{ \langle {#1} | }
\newcommand{\ket}[1]{ | {#1} \rangle }
\newcommand{\EV}{ {\rm eV} }
\newcommand{\KEV}{ {\rm keV} }
\newcommand{\MEV}{ {\rm MeV} }
\newcommand{\GEV}{ {\rm GeV} }
\newcommand{\TEV}{ {\rm TeV} }
\def\diag{\mathop{\rm diag}\nolimits}
\def\Spin{\mathop{\rm Spin}}
\def\SO{\mathop{\rm SO}}
\def\O{\mathop{\rm O}}
\def\SU{\mathop{\rm SU}}
\def\U{\mathop{\rm U}}
\def\Sp{\mathop{\rm Sp}}
\def\SL{\mathop{\rm SL}}
\def\tr{\mathop{\rm tr}}

\def\IJMP{Int.~J.~Mod.~Phys. }
\def\MPL{Mod.~Phys.~Lett. }
\def\NP{Nucl.~Phys. }
\def\PL{Phys.~Lett. }
\def\PR{Phys.~Rev. }
\def\PRL{Phys.~Rev.~Lett. }
\def\PTP{Prog.~Theor.~Phys. }
\def\ZP{Z.~Phys. }

\newcommand{\drawsquare}[2]{\hbox{%
\rule{#2pt}{#1pt}\hskip-#2pt
\rule{#1pt}{#2pt}\hskip-#1pt
\rule[#1pt]{#1pt}{#2pt}}\rule[#1pt]{#2pt}{#2pt}\hskip-#2pt
\rule{#2pt}{#1pt}}

\def\vbr{\vphantom{\sqrt{F_e^i}}}
\newcommand{\fund}{\drawsquare{6.5}{0.4}}
\newcommand{\afund}{\overline{\fund}}
\newcommand{\symm}{\drawsquare{6.5}{0.4}\hskip-0.4pt%
        \drawsquare{6.5}{0.4}}
\newcommand{\asymm}{\raisebox{-3pt}{\drawsquare{6.5}{0.4}\hskip-6.9pt%
        \raisebox{6.5pt}{\drawsquare{6.5}{0.4}}}}
\newcommand{\asymmthree}{\raisebox{-7pt}{\drawsquare{6.5}{0.4}}\hskip-6.9pt%
\raisebox{-0.5pt}{\drawsquare{6.5}{0.4}}\hskip-6.9pt%
\raisebox{6pt}{\drawsquare{6.5}{0.4}}}
\newcommand{\asymmfour}{\raisebox{-10pt}{\drawsquare{6.5}{0.4}}\hskip-6.9pt%
\raisebox{-3.5pt}{\drawsquare{6.5}{0.4}}\hskip-6.9pt%
\raisebox{3pt}{\drawsquare{6.5}{0.4}}\hskip-6.9pt%
        \raisebox{9.5pt}{\drawsquare{6.5}{0.4}}}
\newcommand{\Ythrees}{\raisebox{-.5pt}{\drawsquare{6.5}{0.4}}\hskip-0.4pt%
          \raisebox{-.5pt}{\drawsquare{6.5}{0.4}}\hskip-0.4pt%
          \raisebox{-.5pt}{\drawsquare{6.5}{0.4}}}
\newcommand{\Yfours}{\raisebox{-.5pt}{\drawsquare{6.5}{0.4}}\hskip-0.4pt%
          \raisebox{-.5pt}{\drawsquare{6.5}{0.4}}\hskip-0.4pt%
          \raisebox{-.5pt}{\drawsquare{6.5}{0.4}}\hskip-0.4pt%
          \raisebox{-.5pt}{\drawsquare{6.5}{0.4}}}
\newcommand{\Ythreea}{\raisebox{-3.5pt}{\drawsquare{6.5}{0.4}}\hskip-6.9pt%
        \raisebox{3pt}{\drawsquare{6.5}{0.4}}\hskip-6.9pt
        \raisebox{9.5pt}{\drawsquare{6.5}{0.4}}}
\newcommand{\Yfoura}{\raisebox{-3.5pt}{\drawsquare{6.5}{0.4}}\hskip-6.9pt%
        \raisebox{3pt}{\drawsquare{6.5}{0.4}}\hskip-6.9pt
        \raisebox{9.5pt}{\drawsquare{6.5}{0.4}}\hskip-6.9pt
        \raisebox{16pt}{\drawsquare{6.5}{0.4}}}
\newcommand{\Yadjoint}{\raisebox{-3.5pt}{\drawsquare{6.5}{0.4}}\hskip-6.9pt%
        \raisebox{3pt}{\drawsquare{6.5}{0.4}}\hskip-0.4pt
        \raisebox{3pt}{\drawsquare{6.5}{0.4}}}
\newcommand{\Ysquare}{\raisebox{-3.5pt}{\drawsquare{6.5}{0.4}}\hskip-0.4pt%
        \raisebox{-3.5pt}{\drawsquare{6.5}{0.4}}\hskip-13.4pt%
        \raisebox{3pt}{\drawsquare{6.5}{0.4}}\hskip-0.4pt%
        \raisebox{3pt}{\drawsquare{6.5}{0.4}}}
\newcommand{\Yflavor}{\Yfund + \overline{\Yfund}} 
\newcommand{\Yoneoone}{\raisebox{-3.5pt}{\drawsquare{6.5}{0.4}}\hskip-6.9pt%
        \raisebox{3pt}{\drawsquare{6.5}{0.4}}\hskip-6.9pt%
        \raisebox{9.5pt}{\drawsquare{6.5}{0.4}}\hskip-0.4pt%
        \raisebox{9.5pt}{\drawsquare{6.5}{0.4}}}%

\baselineskip 0.7cm

\begin{titlepage}

\begin{flushright}
SLAC-PUB-12388\\
UT-07-10
\end{flushright}

\vskip 1.35cm
\begin{center}
{\large \bf
Conformal Gauge Mediation}
\vskip 1.2cm
M. Ibe${}^{1}$, Y. Nakayama${}^{2}$ and T.T. Yanagida${}^{2,3}$

\vskip 0.4cm
${}^{1}$ {\it Stanford Linear Accelerator Center, Stanford University,
                Stanford, CA 94309 and} \\
{\it Physics Department, Stanford University, Stanford, CA 94305}\\

${}^2${\it Department of Physics, University of Tokyo,\\
     Tokyo 113-0033, Japan}

${}^3${\it Research Center for the Early Universe, University of Tokyo,\\
     Tokyo 113-0033, Japan}

\vskip 1.5cm

\abstract{We propose a one-parameter theory for gauge mediation of supersymmetry (SUSY)
breaking. The spectrum of SUSY particles such as squarks and sleptons in the SUSY standard-model and the dynamics of SUSY-breaking sector are, in principle, determined only by one parameter in the theory,
that is, the mass of messengers. Above the messenger threshold all gauge coupling and 
Yukawa coupling constants in the SUSY-breaking sector are on the infrared fixed  point. We find that the present theory may predict 
a split spectrum of the standard-model SUSY particles, $m_{\mathrm{gaugino}} < m_{\mathrm{sfermion}}$, where $m_{\mathrm{gaugino}}$ and $m_{\mathrm{sfermion}}$ are SUSY-breaking masses for gauginos and squarks/sleptons, respectively.

}
\end{center}
\end{titlepage}

\setcounter{page}{2}

\section{Introduction}
Gauge mediation \cite{Giudice:1998bp} is an attractive scenario to mediate supersymmetry (SUSY) breaking effects 
from the hidden to the SUSY standard-model (SSM) sector without inducing too large flavor-changing neutral currents. 
This is because the soft SUSY-breaking masses of squarks and sleptons are generated mainly by 
the standard-model gauge interactions and hence they are flavor independent. 
For the successful gauge mediation, we typically introduce messenger chiral superfields 
$P$ and $\bar{P}$ 
which carry standard-model gauge indices with the following superpotential interaction 
to the hidden-sector field $S$:
\begin{eqnarray}
W = S P \bar{P} \ ,
\end{eqnarray}
where the $F$-term of the chiral superfield $\langle S\rangle = m + \theta^2  F_S $ 
breaks the SUSY.

The above minimal gauge mediation model  predicts a definite spectrum of the SSM SUSY 
particles. 
However, the prediction is at least based on two independent parameters $m$ and $F_S$, 
while they are required to conspire together to give a consistent spectrum. 
Besides, in a large class of gauge mediation models, more free parameters are involved, 
and the spectrum of SUSY particles is not uniquely predicted. 
Furthermore, the effective SUSY-breaking parameter $F_S$ might not necessarily correspond 
to the dominant SUSY-breaking scale giving the gravitino mass.

Therefore, one can freely vary the gravitino mass while keeping, for instance, the mass of squarks or gauginos 
in the above gauge mediation models. Indeed, in the gauge mediated SUSY-breaking scenarios, 
an acceptable range of the gravitino mass is restricted only after taking into account the cosmological requirement. 
Then we face a problem from the particle-physics point of view even if the gauge mediation scenario 
is turned out correct: why is the gravitino mass just as we (will) observe in nature? 
Or if the gravitino mass can not be measured: at which scale is the SUSY broken? 

In this paper, we propose ``conformal gauge mediation", where all physical scales in the 
SUSY-breaking sector are fixed by only one parameter, the messenger mass scale. 
Furthermore, since all the coupling constants in the SUSY-breaking sector are fixed by the 
conformal dynamics, there are no arbitrariness between the gravitino mass and gaugino/sfermion  
mass. 
In this sense, our model is comparable to QCD, where relevant physics is 
determined by only one parameter, that is, the dynamical scale $\Lambda_{\mathrm{QCD}}$. 
Similarly, the prediction of our model is determined solely by the 
effective decoupling scale of the messenger particles.

The precise mechanism that makes this construction feasible is provided by a 
(super)conformal invariance of the model at the ultraviolet (UV) cut-off scale 
(e.g. Planck scale). This assumption sets all the free parameters 
appearing in the gauge mediated SUSY-breaking sector at their infrared (IR) fixed-point values. Then, the (super)conformal invariance is broken by one relevant 
parameter of the model, i.e. the mass of the messengers, which triggers the dynamical 
SUSY breaking in the hidden-sector at the same time.

The organization of the paper is as follows. In section 2, we explain the conformal gauge mediation 
scenario by using a concrete and (semi-)calculable example based on the IYIT SUSY-breaking
model. 
The last section is devoted to our conclusions  with some discussion. In appendix, we present technical details for the properties of our extended IYIT model at the conformal fixed point.

\section{Conformal Gauge-Mediation Scenario}
Before going to a concrete example, we explain a basic framework of the conformal 
gauge-mediation scenario.
In our scenario, we consider an extension of a dynamical SUSY-breaking model by adding 
 appropriate number of flavors in vector-like representations
to have an IR fixed point. 

The important assumption we make here is that the above extended SUSY-breaking model is 
in the vicinity of the IR-fixed point at the UV cut-off scale.
Under this assumption, all the coupling constants in the SUSY-breaking sector immediately
converges to the IR-fixed point once they evolve down to the IR from the UV cut-off scale.
Therefore, there remains no free parameter in the conformally extended SUSY-breaking sector 
at the IR scale.

Then, at the far IR scale, the conformal invariance is disturbed by the mass term of the newly introduced flavors
and the SUSY is broken dynamically.
Notice that since all the coupling constants (especially the gauge coupling constant)
of the SUSY-breaking sector is fixed on the IR-fixed point,
the relation between the mass term of the new flavors and the dynamical 
SUSY-breaking scale is uniquely determined. By embedding the SSM gauge group into the flavor symmetry of the new
flavors, we can use them as messenger particles.
In this way, our model has no tunable parameters except for the mass of the
messengers, which eventually determines all the gaugino and the sfermion
masses in the SSM sector.

\subsection{An example of the conformal gauge mediation}

\subsubsection*{Dynamical SUSY-breaking model}

Now, let us consider an example of the conformal gauge mediation based on the IYIT SUSY-breaking model~\cite{IYIT}.
Concretely, we first introduce $Sp(N)$ gauge group with $2(N+1)$ chiral superfields $Q_i$ ($i=1,\dots,2(N+1)$) transforming as a fundamental ($2N$-dimensional) representation of $Sp(N)$. We add singlets $S^{ij} (=-S_{ji})$ in $(N+1)(2N+1)$ representation of the flavor $SU(2N+2)_F$ symmetry of our theory. The superpotential is given by
\begin{eqnarray}
\label{eq:IYIT}
W_{\mathrm{IYIT}} = \lambda S^{ij}Q_iQ_j \ . \label{tree}
\end{eqnarray}

The effective low-energy superpotential is given by
\begin{eqnarray}
W_{\mathrm{eff}} = X(\mathrm{Pf}M_{ij}-\Lambda_{\mathrm{hol}}^{2(N+1)}) + \lambda S^{ij}M_{ij} \label{effIY}
\end{eqnarray}
in terms of gauge invariant low-energy meson superfields $M_{ij} \sim Q_{i}Q_j$. Here $\Lambda_{\mathrm{hol}}$ denotes the holomorphic dynamical scale of our theory \cite{CGT}.
 When $\lambda$ is small, we obtain 
\begin{eqnarray}
\langle M_{ij} \rangle \simeq J_{ij} \Lambda_{\mathrm{hol}}^2 \ , 
\end{eqnarray}
where $J_{ij} = i\sigma_2 \times \mathbf{1}_{N\times N}$ is the invariant tensor of $Sp(N+1)_F \in SU(2N+2)_F$. The effective low-energy dynamics is approximated by the $Sp(N+1)_F$ singlet superfield $S$ with the superpotential
\begin{eqnarray}
W \simeq \lambda \Lambda_{\mathrm{hol}}^2 S \ 
\end{eqnarray}
and the Kahler potential
\begin{eqnarray}
K = S^\dagger S - \frac{\eta}{16\pi^2}\frac{\lambda^4}{\Lambda^2} (S^\dagger S)^2 + \cdots \ ,
\end{eqnarray}
where $\Lambda$ denotes the hadron mass scale of the low-energy effective theory. The $\eta$ is a
constant of order 1 and its signature is not calculable since perturbative calculations break down 
at the hadron mass scale $\Lambda$.

This model generates a dynamical SUSY breaking \cite{IYIT} with\footnote{Strictly speaking, this expression is valid only when $\lambda$ is small. We assume that qualitative features of the SUSY breaking are not modified for large $\lambda$. For example, we have normalized the Kahler potential so that $S$ has a canonical kinetic term. When $\lambda$ becomes large, the definition of our $\Lambda$ might deviate largely from the holomorphic scale $\Lambda_{\mathrm{hol}}$ which is determined by the gauge dynamics alone. This, however, does not affect the main results of our discussion.}
\begin{eqnarray}
F_{S} \simeq \lambda \Lambda_{\mathrm{hol}}^2 \equiv \kappa \lambda \Lambda^2 \ , \label{kappa}
\end{eqnarray}
where we take, in this paper, $ \kappa \simeq 0.1$ as discussed below. 
We, furthermore, assume that $\eta >0$, and hence, the model breaks also $R$-symmetry 
spontaneously as\footnote{When the coupling $\lambda$ is small, we may show that the origin of $S$ is at least a local minimal of the potential \cite{Luty}.}
\begin{eqnarray}
\langle \lambda S \rangle \equiv \Lambda \ ,
\end{eqnarray}
which could be regarded as the definition of our hadron mass scale $\Lambda$.

\subsubsection*{Conformal extension of the SUSY-breaking model}

To extend the IYIT model into the conformal regime, having the gauge mediation in mind, we add messenger chiral superfields $P$ and $\bar{P}$ in the (anti-)fundamental representation  of the standard-model (SM) gauge group (i.e. ${\bf 5}$ and $\bar{{\bf 5}}$ respectively in the grand unified $SU(5)$ gauge group) \cite{Izawa:2005yf} and in the fundamental representations $2N$ under the $Sp(N)$ gauge group. We introduce the SUSY invariant mass term
\begin{eqnarray}
\label{eq:mass}
W_{\mathrm{mass}} = mP\bar{P} \ . \label{mass}
\end{eqnarray}
Well-above the physical mass scale of $P$ and $\bar{P}$, the theory is in the conformal 
regime 
 as we will review momentarily \cite{Ibe:2005pj}. The superpotential \eqref{mass} breaks the conformal invariance and leads to dynamical SUSY breaking as in the original IYIT model below the decoupling scale of $P$ and $\bar{P}$.

So far, the rank of the hidden-sector gauge group has been arbitrary, but we now show $N=2$ is the unique possibility in the conformal gauge mediation scenario. 
Perturbativity of the standard-model gauge interactions demands that the number of the messengers $n_{\mathrm{mess}} (= 2 (N+1))$ should satisfy,
\begin{eqnarray}
n_{\mathrm{mess}}  \lesssim \frac{150}{(1-\gamma_{P})\ln( {M_{\mathrm{GUT}}}/m_{\mathrm{phys}})}\ ,
\end{eqnarray}
where we included the higher loop effects of the $Sp(N)$ sector through the anomalous
dimension $\gamma_{P}$ of $P$ and $\bar{P}$~\cite{NSVZ}.
As we will see shortly, to obtain soft masses of order $100$\,GeV for the SSM SUSY particles, we take 
$m_{\mathrm{phys}} \sim 10^{8}$\,GeV, which yields $n_{\mathrm{mess}} \lesssim 8$ as a bound
for not so strong (i.e., $|\gamma_{P}|\ll1$) hidden sector dynamics.

Thus, when $N>2$, the standard-model gauge couplings will blow up well-below the Planck scale due to the large number of messenger chiral superfields. On the other hand, when $N=1$, the hidden-sector gauge interaction becomes IR-free and we never achieve the IR conformal theory. We, therefore, set $N=2$ in the rest of the paper.

\subsubsection*{Conformal fixed point}
In the conformal gauge mediation scenario, we assume that our SUSY-breaking sector is at the conformal fixed point all the way from the Planck scale down to the SUSY-breaking scale.
Here, we study the properties of the conformal fixed point of the extended IYIT model. 
At the conformal fixed point, all the beta functions of the gauge coupling $\alpha_{\mathrm{hid}}$ of $Sp(2)$ and the Yukawa-coupling $\lambda$ should vanish.

The beta function for the hidden-sector gauge group (for a general $Sp(N)$ gauge group with $n_F$ additional flavors; $N=2$, $n_{F}=5$ in our case) is given by the NSVZ formula~\cite{NSVZ}:
\begin{eqnarray}
 \mu\frac{d}{d\mu} \alpha_{\mathrm{hid}} = 
 -\alpha_{\mathrm{hid}}^2\bigg[\frac{3(N+1)-(N+1)(1-\gamma_Q)-n_F(1-\gamma_{P})}
 {2\pi-(N+1)\alpha_{\mathrm{hid}}}\bigg] \ ,
 \label{eq:betaG1}
\end{eqnarray}
in terms of the anomalous dimensions of the chiral superfields $\gamma_P$ and $\gamma_Q$,
where $\alpha_{\mathrm{hid}}$ is defined in terms of the gauge coupling constant $g_{\mathrm{hid}}$ of $Sp(N)$ as 
$\alpha_{\mathrm{hid}}=g_{\mathrm{hid}}^2/(4\pi)$ and $\mu$ denotes the renormalization scale. The beta function of the Yukawa coupling constant 
in  Eq.(\ref{tree}) is also given in terms of the anomalous
dimension factors of the hyperquarks, $\gamma_Q$, and of the singlet chiral superfields, $\gamma_S$, by
\begin{eqnarray}
 \mu\frac{d}{d\mu} \alpha_\lambda = 
\alpha_\lambda (\gamma_S + 2\gamma_Q) \ ,
\label{eq:betaY1}
\end{eqnarray}
where $\alpha_\lambda$ is defined in terms of the coupling constant $\lambda$ as 
$\alpha_\lambda=\lambda^2/(4\pi)$.  Here and hereafter in the discussion of the conformal fixed point,
we neglect the masses of the messenger quarks $P$ and $\bar{P}$. 

All the beta functions are essentially governed by the wavefunction renormalization factor, or the anomalous dimensions of the chiral fields. The one-loop anomalous dimensions can be computed as\footnote{Here and in the following, we neglect contributions from the standard-model interactions.}
\begin{align}
\gamma_Q &= \frac{2N+1}{2\pi} \alpha_\lambda - \frac{2N+1}{4\pi} \alpha_{\mathrm{hid}} \ ,\label{eq:anomQ1}\\
\gamma_{P}&= - \frac{2N+1}{4\pi} \alpha_{\mathrm{hid}} \ , \label{eq:anomQP1}\\
\gamma_S &= \frac{2N}{2\pi} \alpha_\lambda \ . \label{eq:anomS1}
\end{align}

The requirement for the vanishing beta functions determines the coupling constants at the fixed point as
\begin{align}
\alpha^*_{{\mathrm{hid}}} &= \frac{28\pi}{205} \simeq 0.430 \ , \cr
\alpha^*_{\lambda} &= \frac{2\pi}{41} \simeq 0.153 \ .
\end{align}
The anomalous dimensions of the chiral superfields are computed accordingly as
\begin{align}
\gamma_Q^{\mathrm{1-loop}} &= -\frac{2}{41} \simeq -0.049 \ , \cr
\gamma_P^{\mathrm{1-loop}} &= -\frac{7}{41} \simeq -0.171 \ , \cr
\gamma_S^{\mathrm{1-loop}} &= \frac{4}{41} \simeq 0.096 \ , \ 
\end{align}
in the one-loop (or Banks-Zaks \cite{BZ}) approximation.

Actually, one can obtain the {\it exact} anomalous dimensions by using the $a$-maximization
 technique~\cite{Intriligator:2003jj}
. We only present the final results here (see appendix for the details)
\begin{align}
\gamma_{Q} &= \frac{-211+5\sqrt{1589}}{218} \simeq - 0.0536\ , \cr
\gamma_{P} &= \frac{-1-3\gamma_{Q}}{5}  \simeq -0.168\ , \cr
\gamma_{S} &= -2\gamma_Q \simeq 0.107 \ .
\end{align}
We cannot compute the exact value of the coupling constant at the fixed point from the $a$-maximization technique, but the comparison between the one-loop results and the exact results for the anomalous dimensions suggests the accuracy of about $10\%$.

\subsubsection*{Decoupling and SUSY breaking}
The anomalous dimensions of the chiral fields $P$ and $\bar{P}$  determine the physical decoupling scale $m_{\mathrm{phys}}$ with respect to the mass parameter $m$ at the Planck scale in the superpotential \eqref{tree}. The wavefunction renormalization dictates the relation
\begin{eqnarray}
m_{\mathrm{phys}} = m \left(\frac{m_{\mathrm{phys}}}{M_{Pl}}\right)^{\gamma_P} \ ,
\end{eqnarray}
which leads to
\begin{eqnarray}
m_{\mathrm{phys}} = m \left(\frac{m}{M_{Pl}}\right)^{\frac{\gamma_P}{1-\gamma_P}} \ ,
\end{eqnarray}
where $M_{Pl} \simeq 2.4 \times 10^{18} \mathrm{GeV} $ is the reduced Planck mass.

Let us now discuss relations among various scales.
We note that the physical decoupling scale $m_{\mathrm{phys}}$ and the holomorphic dynamical scale of the IYIT model is related by\footnote{When we use the NSVZ beta function to determine the blow-up scale we obtain $\Lambda_{\mathrm{NSVZ}} \sim 0.13 \times m_{\mathrm{phys}}$.}
\begin{eqnarray}
\Lambda_{\rm hol} = m_{\mathrm{phys}} e^{-\frac{8\pi^2}{g_*^2 b}} = m_{\mathrm{phys}} e^{-\frac{2\pi}{\frac{28\pi}{205}\cdot 6}} \sim 0.087 \times m_{\mathrm{phys}} \ . \label{holo}
\end{eqnarray}
We should note that the hadron mass scale $\Lambda$ does not necessarily coincide with the holomorphic dynamical scale. Later we will set a dynamical assumption $\Lambda \sim 0.3\times m_{\mathrm{phys}}$. Here we consider that the hadron mass scale $\Lambda$ is the scale at which the holomorphic gauge coupling $\alpha_{\rm hol}$ becomes $O(1)$. This means that the parameter $\kappa \simeq 0.1$ in Eq. \eqref{kappa}.

Below the decoupling scale $m_{\mathrm{phys}}$ for $P$ and $\bar{P}$, the dynamics of our conformally extended IYIT model reduces to the original SUSY-breaking IYIT model, albeit the Yukawa coupling $\lambda$ is not so small. 
In this way, we obtain a SUSY-breaking and messenger sector where all the parameters 
are determined only by the messenger mass $m_{\rm phys}$, although it is difficult
to determine them precisely since the interactions becomes strong below the messenger scale and perturbative calculations may break down.

\subsubsection*{Gauge mediation effects}
In the present model, we have not introduced direct couplings between $P$, $\bar P$ and $S$.\footnote{The absence of the direct coupling between the messenger particles and the 
SUSY-breaking fields may  not be a necessary ingredient of the conformal gauge mediation.}
In such a case, the gauge mediation effects to the gaugino masses arise at not 
$O(F_{S}/m_{\rm phys})$ but at $O(F_{S}^{3}/m_{\rm phys}^{5})$~\cite{Arkani-Hamed:1998kj}. Since everything is strongly coupled at the SUSY-breaking scale, 
it is a challenging problem to yield precise values of the soft SUSY-breaking masses for the SSM SUSY particles although 
it should be possible in principle. 
In the following, we give a heuristic evaluation of the soft masses by combining knowledge from perturbative computations and its extrapolation by the naive dimensional analysis, instead of trying to solve this difficult problem in an exact way.

Above the threshold scale of the messengers $P$ and $\bar{P}$, they are in the conformal window. Since the anomalous dimensions of the messengers are not so large, we can integrate out the messengers at the scale $m_{\mathrm{phys}}$, perturbatively. Then, the effective interactions between the SUSY-breaking IYIT sector and the SSM sector are given by, at the leading order, 
\begin{align}
 L_{\mathrm{eff}} & = \frac{\alpha_{\mathrm{SM}}[m_{\mathrm{phys}}]}{4\pi}(4\pi\alpha_{\mathrm{hid}}[m_{\mathrm{phys}}]) \int d^4\theta \frac{\mathrm{Tr}(W_{\mathrm{SM}} W_{\mathrm{SM}})\mathrm{Tr}(W^\dagger_{\mathrm{hid}} W^\dagger_{\mathrm{hid}})}{m^4_{\mathrm{phys}}}
  \cr &+ \frac{\alpha_{\mathrm{SM}}[m_{\mathrm{phys}}]}{4\pi} (4\pi \alpha_{\mathrm{hid}}[m_{\mathrm{phys}}])^2 \int d^4\theta \frac{\mathrm{Tr}(W_{\mathrm{SM}} W_{\mathrm{SM}})\mathrm{Tr}(W^\dagger_{\mathrm{hid}} W^\dagger_{\mathrm{hid}} \bar{D}W^\dagger_{\mathrm{hid}} \bar{D}W^\dagger_{\mathrm{hid}})}{m^8_{\mathrm{phys}}} \cr &+ \cdots  \ , \label{eff}
\end{align}
up to $O(1)$ constants. The SUSY-breaking effects will be mediated to the SSM sector from the above effective interactions.  Notice that the gauge coupling $\alpha_{\mathrm{hid}}[\mu]$ here is not the holomorphic coupling but the canonical one given at the renormalization scale $\mu$.
 The dimension 8 interaction in \eqref{eff} generates soft scalar masses but does not generate gaugino mass,\footnote{This is because perturbative effective interactions between the SSM gauginos and two hidden-sector gauge superfields $W_{\mathrm{hid}}$ in the zero momentum limit of the SSM gauginos are always given by total derivatives. Thus, the gaugino masses vanish.} so we need to keep the next leading dimension 12 interactions.

After the decoupling of messengers $P$ and $\bar{P}$, the conformal invariance is broken and the gauge and Yukawa couplings of IYIT sector run quickly into a nonperturbative regime, leading to a dynamical SUSY-breaking at the scale $\sqrt{F_S} = \sqrt{\kappa \lambda \Lambda^2}$. 
In the following, we present our evaluation of soft masses for the SSM SUSY particles.

After integrating out $Q$ and $\bar{Q}$ (together with the hidden-sector gauge interactions), the SUSY scalar mass is generated by the effective interaction
\begin{eqnarray}
 \left(\frac{\alpha_{\mathrm{SM}}[\Lambda]}{4\pi}\right)^2 \frac{\alpha^2_{\mathrm{hid}}[\Lambda]\lambda^2[\Lambda]}{(4\pi)^2} \int d^4\theta \frac{(\Phi^\dagger\Phi)(S^\dagger S)\Lambda^2}{m^4_{\mathrm{phys}}} \ 
\end{eqnarray}
from the effective three-loop amplitude, where $\Phi$ represents a standard-model superfield. 
This interaction leads to a soft scalar mass
\begin{align}
m_{\rm sfermion} &\simeq \frac{\alpha_{\mathrm{SM}}[\Lambda]}{4\pi} \frac{\alpha_{\mathrm{hid}}[\Lambda]}{4\pi} \lambda[\Lambda] \frac{\sqrt{\langle F_S^\dagger F_S\rangle}\Lambda}{m^2_{\mathrm{phys}}} \cr
&=  \frac{\alpha_{\mathrm{SM}}[\Lambda]}{4\pi}\frac{\alpha_{\mathrm{hid}}[\Lambda]}{4\pi} \lambda^2[\Lambda] \left(\frac{\Lambda}{m_{\mathrm{phys}}}\right)^2 \kappa \Lambda \ , \label{scalar mass}
\end{align} 
where we have used $\langle S \rangle = \Lambda/\lambda(\Lambda)$ and $\langle F_S \rangle = \kappa \lambda(\Lambda) \Lambda^2 $.
Notice that the sign of the sfermion squared masses depends on the sign of the first operator 
in Eq.~(\ref{eff}).\footnote{
This can be seen from the fact that the supertrace of the squared mass matrix of the 
messenger particles $P$ and $\bar P$ is non-vanishing. 
In such cases, the sign of the sfermion masses is sensitive to the details of the
hidden sector~\cite{Poppitz:1996xw}.
 }
However, since the sfermion squared masses also receive  higher-loop corrections after 
integrating out $Q$ $\bar{Q}$ and hidden sector gauge fields, we cannot track the sign
of them perturbatively. 
In the following, we simply assume that the sign of the squared mass of sfermions are positive.

On the other hand, we obtain the effective interaction for gauginos~\cite{Arkani-Hamed:1998kj},
\begin{eqnarray}
 \frac{\alpha_{\mathrm{SM}}[\Lambda]}{4\pi} \frac{\alpha^4_{\mathrm{hid}}[\Lambda]\lambda^4[\Lambda]}{(4\pi)^4} \int d^4\theta \frac{\mathrm{Tr}(W_{\mathrm{SM}}W_{\mathrm{SM}})(S^\dagger S)(S^\dagger D^2 S)\Lambda^2}{m^8_{\mathrm{phys}}} \ 
\end{eqnarray}
from the effective four-loop diagrams with the use of the dimension 12 operators in \eqref{eff}.\footnote{Originally before integrating out messengers, they correspond to five-loop diagrams.} Note that we have used the hidden-sector gauge coupling and the Yukawa coupling at the hadron mass scale $\Lambda$, where the SUSY-breaking effects will dominate. 
This interaction leads to a soft gaugino mass 
\begin{align}
m_{\rm gaugino} &\simeq \frac{\alpha_{\mathrm{SM}}[\Lambda]\alpha^4_{\mathrm{hid}}[\Lambda]}{(4\pi)^5} \lambda^4[\Lambda] \frac{ \langle S^\dagger \rangle \langle F_S\rangle\langle F^\dagger_S\rangle\langle F_S\rangle \Lambda^2}{m^8_{\mathrm{phys}}} \cr
&= \frac{\alpha_{\mathrm{SM}}[\Lambda]\alpha^4_{\mathrm{hid}}[\Lambda]}{(4\pi)^5} \lambda^6[\Lambda] \left(\frac{\Lambda}{m_{\mathrm{phys}}} \right)^8 \kappa^3 \Lambda \ . \label{gaugino mass}
\end{align}
We note that the contribution is suppressed compared with the expression proposed in \cite{Izawa:2005yf} in the context of the strongly coupled gauge mediation, by the factor of $|F_S|^2/{m^4_{\mathrm{phys}}}$.\footnote{As discussed in \cite{Arkani-Hamed:1998kj}, the naive contribution to the gaugino mass $m_{{1}/{2}} \sim \alpha_{\mathrm{SM}}\alpha_{\mathrm{hid}}^2 F_S \langle \lambda S\rangle/{m^2_{\mathrm{phys}}}$ should be suppressed either by $|F_S|^2/{m^4_{\mathrm{phys}}}$ or the standard-model loop factor $\alpha_{\mathrm{SM}}^2$ from the gaugino screening mechanism. }

{}From \eqref{scalar mass} and \eqref{gaugino mass}, we obtain a relation between the gaugino and scalar masses as
\begin{align}
m_{\rm gaugino} &= \left(\frac{\alpha_{\mathrm{hid}}[\Lambda]}{4\pi}\right)^3 \lambda^2[\Lambda] \frac{|F_S|^2\Lambda^2}{m^6_{\mathrm{phys}}} m_{\rm sfermion} \cr
&= \left(\frac{\alpha_{\mathrm{hid}}[\Lambda]}{4\pi} \right)^3 \lambda^4[\Lambda] \kappa^2\left(\frac{ \Lambda}{m_{\mathrm{phys}}}\right)^6 m_{\rm sfermion}\ . \label{ratio}
\end{align}

\subsubsection*{Evaluation of soft masses}
Let us go on to estimate the soft masses of the SSM SUSY particles.
Now we postulate some dynamical assumptions of our model from the naive dimensional analysis and intuitions from other strongly coupled field theories such as QCD. From the naive dimensional analysis, we naturally assume the canonical gauge coupling is given by $\alpha_{\mathrm{hid}}[\Lambda] \sim 4\pi.$\footnote{
The holomorphic gauge coupling 
$\alpha_{\rm hol}$ is of $O(1)$ at the hadron mass scale $\Lambda$. However, the canonical coupling $\alpha_{\mathrm{hid}}$ may become much larger than $\alpha_{\rm hol}$ there, since it
receives higher-order corrections.}

In addition, there are several uncalculable parameters in our model: $\Lambda/m_{\mathrm{phys}}$, $\lambda[\Lambda]$ and $\kappa$. First of all, the prediction of our model depends sensitively on the ratio $\Lambda/m_{\mathrm{phys}}$. As explained in the previous discussion, we take $(\Lambda/m_{\mathrm{phys}}) \sim 0.3 $.\footnote{The one-loop holomorphic dynamical scale $\Lambda_{\mathrm{hol}}$ has a relation $\Lambda_{\mathrm{hol}} \sim 0.1 \times m_{\mathrm{phys}}$ as in \eqref{holo}. However, the higher loop corrections may render the hierarchy between $\Lambda$ and $m_{\mathrm{phys}}$ smaller. }

We suppose $ \alpha_\lambda[\Lambda] \simeq 1 $ since the Yukawa coupling tends to be smaller than the  gauge coupling constant in the perturbative regime (but non-perturbatively large at the hadron mass scale $\Lambda$). We also recall that the above assumption on $(\Lambda/m_{\mathrm{phys}}) \sim 0.3$ corresponds to the parameter $\kappa\simeq 0.1$, since $\Lambda_{\mathrm{hol}} \sim 0.1 \times m_{\mathrm{phys}}$. Note that our results for gaugino and scalar masses depend only on a particular combination of $\kappa$ and $\lambda[\Lambda]$, namely $\kappa \lambda^2[\Lambda]$. Thus our dynamical assumptions above amount to
\begin{eqnarray}
\kappa \lambda^2[\Lambda] \simeq 1 \ .
\end{eqnarray}

Then, the ratio between the gaugino and the sfermion masses becomes
\begin{eqnarray}
\frac{m_{\rm gaugino}}{m_{\rm sfermion}} \sim 10^{-3} \ .
\end{eqnarray}
Our model, therefore, predicts a split spectrum of the SSM SUSY particles. 
Under the same assumptions, we have gaugino masses of order $100$\,GeV for $\Lambda \simeq 10^{8}$\,GeV. It also determines soft SUSY-breaking masses $m_{\mathrm{sfermion}}$ for squarks and sleptons and the gravitino mass as $m_{\rm sfermion} \sim 100$ TeV and $m_{3/2} \sim 1$\,MeV, where we have used $m_{3/2} = {\kappa \lambda \Lambda^2}/{(\sqrt{3}M_{Pl})}$. 
 
\subsubsection*{Comments on the model}
Before we finish our discussion on the present model, several comments are in order.
Firstly, notice that the effective coupling of the messengers to the SUSY-breaking fields
is governed by the $Sp(2)$ gauge interaction, and hence it is independent of the charges of the messenger particles under the SSM gauge group.
On the other hand, the mass of the colored messengers becomes heavier than the one of the 
$SU(2)_{L}$-doublet messengers because of the difference of the wavefunction renormalization
of the messenger particles due to the SSM gauge interactions.
As a result, the gluino mass to the wino mass ratio is suppressed by a factor of,
\begin{eqnarray}
\frac{m_{\rm gluino}}{m_{\rm wino}}\sim \left(\frac{\alpha_{3}}{\alpha_{2}}\right)\times
\left(\frac{m_{\rm phys}^{SU(2)_{L}}}{m_{\rm phys}^{SU(3)_{c}} }
\right)^{8},
\end{eqnarray}
where $\alpha_{2,3}$ denote the coupling constant of $SU(2)_{L}$ and $SU(3)_{c}$,
and $m_{\rm phys}^{SU(3)_{c},SU(2)_{L}}$ the messenger masses of the colored
and the $SU(2)_{L}$-doublet ones, respectively.
Then, assuming $m_{\rm phys}\sim 10^{8}$\,GeV and $m_{\rm phys}^{SU(2)_L}=m_{\rm phys}^{SU(3)_c}$ at the GUT scale, we find that the mass ratio is 
given by $m_{\rm phys}^{SU(3)_{c}} \simeq 1.2 \times m_{\rm phys}^{SU(2)_{L}}$, which leads to 
$m_{\rm gluino}/m_{\rm wino}\sim 0.8$.
Therefore, our model predicts the gluino lighter than the wino.

Secondly, we point out that our model as it is may suffer from a problem associated 
with the Nambu-Goldstone bosons. 
If our theory is on the IR fixed point at the Planck scale, we have an exact global $SU(6)$ symmetry in the IYIT sector. 
Thus, we have a number of Nambu-Goldstone bosons in the SUSY-breaking vacuum. 
To eliminate such massless bosons we consider a milder assumption that our theory is near the fixed 
point at the Planck scale and hence explicit breakings of the global $SU(6)$ symmetry remain in the Yukawa coupling $\lambda$ at the SUSY-breaking scale.

Once we accept some small deviations from the conformal theory, it may be natural to consider possible tiny Yukawa couplings of the $S$ field to the massive quarks $P$ and ${\bar P}$.
Interesting is that such a coupling $\lambda 'SP{\bar P}$ induces a gaugino mass at the one-loop level and hence the gaugino mass is no longer suppressed. We find a mild hierarchy among soft SUSY-breaking masses as $m_{\rm gaugino}\simeq O(0.1) \times m_{\rm sfermion}$ for $\alpha_{\lambda '}\simeq 10^{-3}$,
for instance.\footnote{Here, we are treating $\lambda'$ as a small perturbation to the
conformal theory of $\alpha_{\mathrm{hid}}$ and $\lambda$.
However, if there is another fixed point with $\lambda' \neq 0$, we can consider 
another candidate of the conformal gauge mediation around such a new fixed point~\cite{INY}. For such a conformal model, the ratio between 
the $m_{\rm gaugino}$ and $m_{\rm sfermion}$ becomes the same as in the conventional
gauge mediation models; $m_{\rm gaugino}/m_{\rm sfermion}\simeq \sqrt{n_{\rm mess}}$.
}
 In this case, the gravitino mass can be as low as $m_{3/2}\simeq O(10)$ eV, 
provided $m_{\rm gaugino} \simeq O(100)$\,GeV and $m_{\rm sfermion} \simeq O(1)$\,TeV. Then, the model possesses no cosmological problems \cite{Viel:2005qj}, but
on the other hand, we lose a candidate for the dark matter. It should be also noted that the new Yukawa coupling $\lambda 'SP{\bar P}$ generates a SUSY-preserving true vacuum at $S=m_{\mathrm{phys}}/\lambda '$. We find that the tunneling probability from our SUSY-breaking vacuum to the true vacuum is strongly suppressed for the above small Yukawa coupling $\lambda '$.

\subsection{Other examples}
The large disparity between $m_{\rm gaugino}$ and $m_{\rm sfermion}$ 
in the above model is originated from the smallness of the ratio,
${\Lambda}/{m_{\mathrm{phys}}}\simeq 0.3$. 
However, if the fixed-point value of the gauge coupling constant is of order 1, 
the ratio of ${\Lambda}/{m_{\mathrm{phys}}}$ may become $O(1)$ and we may have
a much milder  disparity of the SUSY spectrum without a direct coupling 
between messenger particles and the SUSY-breaking field.
For example, let us consider the (uncalculable) dynamical SUSY-breaking
model of $SU(5)$ with $\mathbf{10}+\bar{\mathbf{5}}$~\cite{Affleck:1983vc}. 
We can add $N_f$ vector-like quarks $\mathbf{5} +
\bar{\mathbf{5}}$ to make it a conformal field theory (for $ 5<N_f <13$).\footnote{A similar model has been studied in \cite{Ibe:2005qv} (see also \cite{Pouliot:1995me}). The detailed analysis will be given elsewhere~\cite{INY}.}
We assume that five 
pairs of additional $N_f$ quarks are charged under the standard-model gauge group and they serve as
messengers. The anomalous dimensions of the chiral superfields at the
conformal fixed point can be computed by using the $a$-maximization
technique:
\begin{align}
\gamma_{\mathbf{5}} &= \gamma_{\bar{\mathbf{5}}} =
\frac{-85+8(-14+N_f)N_{f}+3\sqrt{5425-8N_{f}(1+N_f)}}{-25+8N_{f}(1+N_f)} \ , \cr
\gamma_{\mathbf{10}}&= \frac{1}{3}(-26+2N_f-2N_f \gamma_{\mathbf{5}} - \gamma_{\mathbf{5}}) \ .
\end{align}
The gauge coupling constant at the fixed point can be computed as
\begin{eqnarray}
\alpha_* \simeq -\frac{5\pi}{12} \gamma_{\mathbf{5}}
\end{eqnarray}
in the one-loop approximation. For $N_f = 6$, we have $\alpha_* \sim 1.1$.
Under the assumption that all the messengers decouple at the same scale $m_{\mathrm{phys}}$,
this model may provide a mild hierarchy between the gauginos and sfermions
because the ratio of ${\Lambda}/{m_{\mathrm{phys}}}$ may be $O(1)$.

\section{Conclusions}
In this paper, we have proposed a concept of the conformal gauge mediation. The conformal invariance at the cut-off scale removes free parameters in the conventional gauge mediations. As a concrete example, we have constructed a conformal theory based on the extension of the IYIT model. Although the requirement of the conformal invariance, the dynamical SUSY-breaking and the perturbativity of the standard-model gauge interactions severely restricts a viable class of models for the conformal gauge mediation, it should be possible to construct other examples based on different dynamical (possibly metastable) SUSY-breaking models \cite{Kitano:2006xg}.

A crucial point in the conformal gauge mediation scenario is that there is a strict relation between 
SUSY-breaking masses for the SSM SUSY particles and the gravitino mass. 
 If the gravitino 
is the dark matter in the universe, it should have a mass, $ m_{3/2} \gtrsim 550$\,eV, 
to be sufficiently cold as required from astrophysical observations \cite{Viel:2005qj}.
This bound is transferred to the lower bound on the squark masses as $m_{\rm sfermion} \ge O(1)$ TeV in the conformally extended SUSY-breaking IYIT model. 
This might explain the null result of the Higgs boson search for $m_{\rm higgs}\lsim 114$\,GeV.

\section*{Acknowledgements}
Y.N. thanks the Japan Society for the Promotion of Science for financial
support. 
M.I. thanks R.~Kitano for useful discussions.
This work was supported by the U.S. Department of Energy under contract
number DE-AC02-76SF00515.

\appendix
\section{Anomalous dimensions from $a$-maximization}
In this appendix we use the so-called $a$-maximization method~\cite{Intriligator:2003jj}
to determine the anomalous dimensions of the fields
in the conformally extended IYIT model
beyond the Banks-Zaks approximation presented in section 3.

The $a$-maximization method simply states that the conformal $R$ current
appearing in the superconformal algebra maximizes
a particular 't Hooft anomaly
\begin{equation}
a = \mathrm{Tr}(3R^3 - R) \ ,
\end{equation}
which is related to the conformal anomaly on a curved spacetime
\begin{eqnarray}
\int_{S^4} \langle T_\mu^\mu\rangle \ .
\end{eqnarray}

In our model of the $Sp(N)$ gauge theory,
the candidate of the conformal $R$ current
contains one free parameter $x=\gamma_Q$,
from which the corresponding $R$ charges
are determined as
\begin{eqnarray}
R_Q = \frac{2}{3}(1+\frac{x}{2})\ , \quad
R_{P} = \frac{2}{3}(1+\frac{\gamma_{P}}{2})\ , \quad
R_{S} = \frac{2}{3}(1+\frac{-2x}{2})
\end{eqnarray}
with
\begin{equation}
\gamma_{P} = 1-\frac{3(N+1)-(N+1)(1-x)}{2(N+1)-\varepsilon}\ ,
\end{equation}
where $\varepsilon = 2(N+1)-n_F$. 

The claim is that among these one-parameter $R$ currents,
the conformal one maximizes the anomaly $a$,
which is obtained
as follows:
\begin{align}
a &= 2N(2N+2)\left[3(R_Q-1)^3-(R_Q-1)\right] \cr
  &  +2(2N+2-\varepsilon)2N \left[3(R_{P}-1)^3-(R_{P}-1)\right] \cr
  &  + (N+1)(2N+1)\left[3(R_S-1)^3-(R_S-1)\right] \ ,
\label{a}
\end{align}
where we note that the $R$ charges appearing in $a$ are those of fermions
(i.e. $R_{\psi_{Q}} = R_{Q}-1$) because only fermions contribute to the anomaly.
By maximizing $a$ with respect to $x$, we can determine $x_{*}=\gamma_{Q}|_*$. The unique local maximum is achieved by setting 
\begin{align}
x_* &= -\left(\varepsilon^2(2+3N)-4\varepsilon(1+N)(2+3N)
+(1+N)^2(8+13N)\right)^{-1} A \ ; \cr
A &\equiv 4-4\varepsilon+\varepsilon^2 + 22N-16\varepsilon N
 + 3 \varepsilon^2 N + 32N^2 - 12 \varepsilon N^2 + 14 N^3 
 + (\varepsilon - 2(1+N))B \ , \cr
B &\equiv \sqrt{\varepsilon^2(1+2N)(1+6N)
-4\varepsilon (1+N)(1+2N)(1+6N)+ (2+9N+7N^2)^2}\ .
 \label{horrible}
\end{align}

\end{document}